\documentclass[runningheads]{llncs}
\usepackage{graphicx}%
\usepackage{multirow}%
\usepackage{amsmath,amssymb,amsfonts}%
\usepackage{mathrsfs}%
\usepackage[title]{appendix}%
\usepackage{xcolor}%
\usepackage{textcomp}%
\usepackage{manyfoot}%
\usepackage{booktabs}%
\usepackage{algorithm}%
\usepackage{algorithmicx}%
\usepackage{algpseudocode}%
\usepackage{listings}%
\usepackage{dirtytalk}%
\usepackage{csquotes}
\usepackage{todonotes}
\usepackage{xcolor}
\usepackage{comment}
\usepackage{algpseudocode}
\usepackage{algorithm}
\usepackage{pgfplots}
\usetikzlibrary{patterns}
\usepackage{caption}
\usepackage{float}
\usepackage{cite}
\usepackage{subcaption}
\pgfplotsset{compat=1.18}

\usepackage[hidelinks]{hyperref}
\hypersetup{
    colorlinks=true,
    linkcolor=blue,
    filecolor=blue,      
    urlcolor=blue,
}

\begin{document}
\title{Hook-in Privacy Techniques for \\gRPC-based Microservice Communication}

\author{
Louis~Loechel \orcidID{0000-0002-5877-3706} \and
Siar-Remzi~Akbayin \orcidID{0009-0009-3415-0449} \and
Elias~Grünewald \orcidID{0000-0001-9076-9240} \and
Jannis Kiesel \orcidID{0000-0002-7412-3746} \and
Inga~Strelnikova \orcidID{0009-0003-0980-0944} \and
Thomas~Janke \orcidID{0009-0009-6021-9817} \and
Frank~Pallas \orcidID{0000-0002-5543-0265}
}
\authorrunning{Loechel et al.}
\institute{Information Systems Engineering, Technische Universität Berlin}
\maketitle              %
\begin{abstract}

gRPC is at the heart of modern distributed system architectures. Based on HTTP/2 and Protocol Buffers, it provides highly performant, standardized, and polyglot communication across loosely coupled microservices and is increasingly preferred over REST- or GraphQL-based service APIs in practice. Despite its widespread adoption, gRPC lacks any advanced privacy techniques beyond transport encryption and basic token-based authentication. Such advanced techniques are, however, increasingly important for fulfilling regulatory requirements. For instance, anonymizing or otherwise minimizing (personal) data before responding to requests, or pre-processing data based on the purpose of the access may be crucial in certain usecases. In this paper, we therefore propose a novel approach for integrating such advanced privacy techniques into the gRPC framework in a practically viable way. Specifically, we present a general approach along with a working prototype that implements privacy techniques, such as data minimization and purpose limitation, in a configurable, extensible, and gRPC-native way utilizing a gRPC interceptor. We also showcase how to integrate this contribution into a realistic example of a food delivery use case. Alongside these implementations, a preliminary performance evaluation shows practical applicability with reasonable overheads. Altogether, we present a viable solution for integrating advanced privacy techniques into real-world gRPC-based microservice architectures, thereby facilitating regulatory compliance ``by design''.

\keywords{gRPC \and Microservices \and Privacy \and Purpose Limitation \and Data Minimization \and API \and Cloud Native \and Web Engineering.} %
\end{abstract}

\section{Introduction}
\label{intro}
Microservice architectures, 
which divide a system into many small services that all fulfill a specific business capability or %
purpose, %
have established as the prevailing paradigm for implementing and operating complex, large-scale web systems and applications \cite{nadareishvili2016microservice}.
In cloud native computing environments, %
respective microservices
materialize as containerized, loosely-coupled system components \cite{microserv}. Meanwhile, agile development teams and DevOps practices further support the use of different technology stacks (incl. programming languages) per microservice and, consequently, allow separating teams accordingly \cite{jabbari2016devops}. However, to leverage these advantages, microservice architectures need language-agnostic or at least polyglot interfaces such as Representational State Transfer (REST), GraphQL, or Remote Procedure Calls (RPC) which enable efficient communication between different services. The utilization of a specific Application Programming Interface (API) paradigm depends on the usecase and, e.g., the data characteristics (cf.~Sect.~\ref{section:background}).

Alongside these technical developments towards microservice architectures, the importance of privacy regulations -- such as the GDPR \cite{GDPR}, the CCPA and others -- and the need to properly address them technically (\say{by design}) is increasingly recognized. Noteworthily, this goes way beyond mere access restrictions but calls for nuanced measures: The privacy principle of \emph{data minimization} (embodied in, e.g., Art.~5(1c) of the GDPR), for instance, requires that personal data are \say{limited to what is necessary in relation to the purposes for which they are processed}. The principle of \emph{purpose limitation} (Art.~5(1(b)), in turn, requires that personal data are only used for those purposes they were originally collected for (or for those purposes deemed compatible with the initial ones). One and the same data-providing service must therefore respond with different \say{views} to the same data, depending on the access context \cite{data-min-pallas}. Further privacy principles induce similar or additional needs, but we exemplarily confine ourselves to these two herein.

With
large microservice architectures %
consisting of hundreds of services -- using different technology stacks \cite{microserv} and independently developed and maintained by different teams -- %
adherence to such requirements cannot be achieved by manual implementation or audit. %
Instead, compliance must be supported through configurable %
technical approaches, which implement privacy principles on a per-service basis. %
To date, however, developers %
lack the means to do so \cite{devprivops}. %
In particular,
API frameworks, such as gRPC,\footnote{\href{https://grpc.io/docs/what-is-grpc/faq}{grpc.io/docs/what-is-grpc/faq}} expose an inherent lack of advanced privacy techniques that go beyond mere transport encryption and simple token-based authentication. Such advanced techniques are, however, indispensable for properly addressing said principles. So far, developers can thus either go without appropriate technical implementation of privacy requirements within their services (leaving compliance to rather non-technical means) or implement required techniques manually, in a rather ad-hoc fashion (raising excessive efforts as well as the risk of errors and improper implementations).

First proposals to close this gap have been made for services exposed via GraphQL \cite{data-min-pallas}, but for the whole field of performance-sensitive microservices communicating via gRPC, the need to integrate advanced privacy techniques in a configurable and performance-aware manner remains largely untapped. In consequence, we herein propose and contribute:

\begin{itemize}
    \item A general approach for \textbf{hook-in %
    privacy techniques in high-performance remote procedure call frameworks}, especially applicable in cloud native microservices,
    \item a \textbf{proof-of-concept implementation} of our approach for the widely-used, enterprise-grade
    gRPC~framework in a polyglot, cloud-native microservice environment, exemplified through the privacy principles of data minimization and purpose limitation, and 
    \item a preliminary \textbf{performance evaluation} in a realistic food delivery scenario.
\end{itemize}

These contributions unfold as follows:
Background and related work are explained in Sect.~\ref{section:background}.
In Sect.~\ref{section:requirements}, we identify the requirements to be fulfilled. Our %
general approach is presented in Sect.~\ref{section:approach}, followed by our proof-of-concept implementation in Sect.~\ref{section:implementation} %
and a preliminary performance evaluation in Sect.~\ref{section:evaluation}. Sect.~\ref{section:conclusion} discusses our results, identifies %
prospects for future work, and concludes.

\section{Background and Related Work}
\label{section:background}

Our work builds on the following foundations and related work. %

\subsection{Microservices Communication via gRPC}

One popular communication method between microservices is the Remote Procedure Call (RPC). RPCs are a way to invoke procedures across machines, while it looks like a single-machine execution from a developer's perspective \cite{RPC_Definition,rpc70}. %

One of the most popular RPC frameworks, gRPC, was initially developed internally at Google and open-sourced in 2015.\footnote{\href{https://developers.googleblog.com/2015/02/introducing-grpc-new-open-source-http2.html}{developers.googleblog.com/2015/02/introducing-grpc-new-open-source-http2.html}} It is an efficient and scalable framework for inter-service communication %
implementing RPCs over HTTP/2 \cite{brown2023measuring}. 
Furthermore, it supports basic authentication mechanisms, streaming, blocking / non-blocking transmission, etc., and is available for a broad variety of programming languages. 
By default, it uses Protocol Buffers\footnote{\href{https://protobuf.dev/overview}{protobuf.dev/overview}} for serializing structured data in a forward- and backward-compatible way.
Protocol Buffers support many languages by default %
and even more through third-party add-ons \cite{kumar2021performance}. %
The definition of the data structure has to be defined in a \texttt{.proto} file which is then used by the \texttt{protoc} compiler to generate the necessary code in the chosen language which can then be used by the application \cite{kumar2021performance}.

Using gRPC is most suitable for communication between microservices in a cloud environment, while, for the browser interface, alternatives such as REST or GraphQL are the preferred options \cite{kumar2021performance}. Thus far, privacy-enhancing technologies, including data minimization and purpose limitation, are mostly lacking \cite{agape2018charting}. %

\subsection{Technical Approaches for Privacy Techniques in Inter-Service Communication}

In related work, an approach on how to implement purpose-based access control on the application layer is proposed \cite{pallas2020towards}. Their work presents two prototype implementations %
with respective benchmarks. %
This informs our work regarding ease of implementation.
Furthermore, the \textsc{Janus} prototype provides a viable approach, which extends the popular Apollo server to introduce attribute-level access control and role-based data minimization mechanisms to GraphQL APIs \cite{data-min-pallas}. \textsc{Janus} employs JSON Web Tokens (JWTs) to identify roles and, on this basis, parameterize the application of  %
common data minimization techniques in a per-request fashion. With its flexible hook-in capabilities, \textsc{Janus} shall thus serve as a blueprint for our endeavor to implement similar capabilities into gRPC-based service APIs.%

Specifically related to gRPC and Protocol Buffers, previous work proposes the implementation of data flow assertions \cite{brown}.
Put briefly, a Go library here generates access policies based on JSON files and a gRPC interceptor inspects the HTTP request headers. Therefore, access control is purely based on the encryption of strings. The interceptor only decrypts the data after comparing the headers with the policies. Encrypting every string in a message by default makes the overhead of this approach not feasible for high-performance scenarios. 

Beyond this work, gRPC interceptors are used for security (authentication),\footnote{\href{https://grpc.io/docs/guides/auth/}{grpc.io/docs/guides/auth}} observability practices (tracing),\footnote{\href{https://pkg.go.dev/go.opentelemetry.io/contrib/instrumentation/google.golang.org/grpc/otelgrpc}{go.opentelemetry.io/contrib/instrumentation/google.golang.org/grpc/otelgrpc}} or fault tolerance mechanisms (retries).\footnote{\href{https://pkg.go.dev/github.com/grpc-ecosystem/go-grpc-middleware/v2/interceptors/retry}{pkg.go.dev/github.com/grpc-ecosystem/go-grpc-middleware/v2/interceptors/retry}} Approaches utilizing the interceptor concept to implement privacy techniques such as data minimization and purpose limitation in high-performance settings are, however, to the best of our knowledge not existing.

\subsection{Data Minimization and Purpose Limitation in Inter-Service Communication} %
Established privacy-preserving techniques, such as \textit{suppression}, \textit{generalization} and \textit{noising}, serve as a foundation for this work regarding data minimization \cite{sweeney2002k, majeed_anonym, marques2020analysis}. 
Likewise, purpose limitation techniques following the idea of the Purpose-Based Access Control model are incorporated prototypically into this work \cite{byun2005purpose, byun2008purpose, pallas2020towards}.
Furthermore, we build upon the field of access control. The eXtensible Access Control Markup Language (XACML) \cite{xacmlOriginal} and its respective component model have been widely adopted as a standard for creating fine-grained policy rules \cite{xacmlSurvey,xacmlIot}. Within the XACML framework, access control operations are partitioned into distinct functional components, which are the Policy Administration Point (PAP), Policy Decision Point (PDP), and Policy Enforcement Point (PEP). 
This component model is frequently applied to the privacy principles of purpose limitation \cite{byun2005purpose}, data minimization, and their overlaps with traditional technical access control measures \cite{chandramouli2021attribute,finck2021reviving,biega2020operationalizing}.%

\section{Requirements}
\label{section:requirements}
In line with other privacy engineering endeavors (such as \cite{pallas2020towards, data-min-pallas, hawk}) we outline a set of reasonable functional and non-functional requirements. %

\textit{Policy-Based Data Minimization (FR1):} Derived from the privacy principle of data minimization (codified, e.g., in Art.~5 GDPR) access to specific personal information must be restricted as far as possible. In particular, it may not always be necessary to expose the complete set of accessed information. The proposed solution should therefore apply different types of data minimization mechanisms. These include noising, suppression, or advanced anonymization techniques \cite{biega2020operationalizing}.

\textit{Policy-Based Purpose Limitation (FR2):} For enabling basic purpose limitation \cite{wolf2021messaging}, it is required that each category of personal data and every gRPC call can be supplemented with specific processing purposes. This allows the utilization of various purpose-based access policies. %
Related to Art.~5 GDPR, this provides the means to specifically control which calls and services can access personal data.

\textit{Configurability (FR3):} The proposed solution must be highly configurable to facilitate adoption in different use case scenarios. Developers must therefore be able to specify and choose a domain-specific set of available purposes, according to their system-specific needs. This is desirable because the required processing purposes of different domains can diverge greatly. 

\textit{Native gRPC integration (FR4):} As mentioned in Sect.~\ref{section:background}, many technologies can map gRPC to be interoperable with different communication protocols and data structures, which are not gRPC-native. However, whenever possible the solution should consist of gRPC-native protocols and data structures to keep its benefits such as high performance as well as low integration overhead, and without adding another level of complexity.

\textit{Reasonable Performance Overhead (NFR1):} According to Art.~25 GDPR, technical measures that realize privacy principles must be applied while taking into account the cost of implementation, which includes induced overheads. We will therefore assess the performance of our approach with real-world settings and configurations to ensure that the overhead is at a reasonable level.

\textit{Polyglot compatibility (NFR2):} To ensure that our approach is compatible with a wide range of programming languages and frameworks, it has to be implemented in a modular and extensible manner. This will allow developers to easily integrate the privacy-enhancing technologies into their existing applications. Additionally, clear documentation and examples shall guide developers in the integration process.

\section{Approach}\label{section:approach}

To fulfill the functional and non-functional requirements including regulatory obligations, we propose a conceptual approach for effective and efficient data minimization and purpose limitation for high-performance gRPC-based inter-service communication in microservice architectures.

Derived from \textit{FR1}, \textit{FR2} as well as \textit{FR4}, our general approach for integrating privacy techniques, such as purpose limitation and data minimization, will be realized as a server-side middleware. In our implementation, we opt for adding a gRPC response interceptor. 
This allows the integration in the required gRPC-native manner, while working on an abstraction layer that does not require existing microservices and their respective code base to be modified with more than about two lines of code \textit{(NFR2)}. Additionally, to meet the performance requirements outlined in \textit{NFR1}, we minimize the complexity of the interceptor. Therefore, any computation that is not required to be performed immediately with each response will be conducted in a separate system component, independent of the interceptor itself. 
Following the XACML component model, access policy enforcement is preceded by policy administration and interpretation in context-specific scenarios. Thus, to allow for a scalable interceptor, the PAP and the PDP are consciously separated from the PEP.

\begin{figure}[!t]
    \centering
    \includegraphics[width=0.9\textwidth]{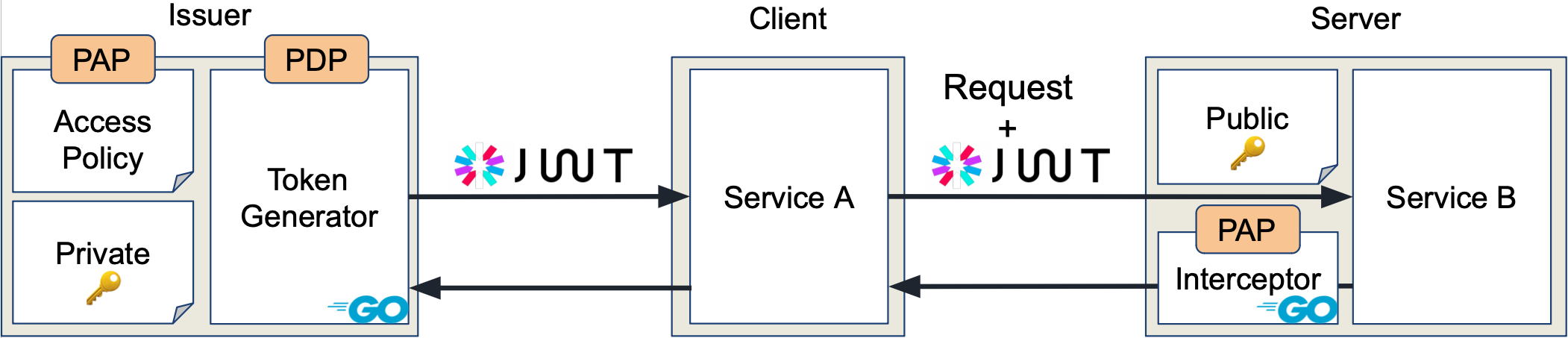}
    \caption{Architectural overview representing the communication process between client and server using JWT in gRPC communication incl. the XACML-inspired control functionality mapping.}
    \label{fig:contribution}
\end{figure}

The separation of the PAP and PDP from the PEP not only reduces the anticipated performance overhead from the interceptor but also necessitates the establishment of reliable and efficient communication for decisions made at the PDP. We address this challenge by implementing signed JSON Web Tokens (JWTs). A JWT typically consists of three parts: header, payload, and signature. The payload consists of claims that can be exchanged between different parties securely \cite{jwt}. These tokens will serve as a trusted certificate enabling parties to exchange decisions made at the PDP, prior to the PEP. Thus, a client, Service A, intending to send a gRPC request to a server, Service B, must first get such a trusted JWT as illustrated in Fig.~\ref{fig:contribution}. We propose to sign the message via standard asymmetric cryptography, including a public/private key pair.

To reduce the computational and communicative overhead, a given JWT remains valid for a specified time. This method assumes that neither the access policy nor the context on which the decision is based are subject to frequent alterations. Therefore, neither the PAP nor the PDP require round-trips for each outgoing response. After validating the JWT, the interceptor merely executes the policy decision upon receiving an outgoing server response. For more details, we provide an in-depth explanation of the JWT implementation in Sect.~\ref{PAP}.

Assuming the client possesses a valid JWT containing a recent access policy decision, it can proceed to communicate with other microservices within the architecture. The JWT is appended to the outgoing context of the message whenever a request is sent to another microservice. Other data stored within the context are not affected by this addition, nor does it prevent future alterations, which substantially enhances integration.
Upon request arrival at the server, the procedure is handled regularly. Data is aggregated to form the response message if the request invokes a response. The interceptor acts as the policy enforcing middleware as soon as the server sends the response back to the client.

gRPC interceptors can generally interact with almost all facets of a message, including the payload, which may contain personal data. For applying privacy techniques, the message payload is central to modifications. Responses sent via gRPC can contain multiple data fields, each comprising a field name and its value. An accompanying access policy, retrieved from the JWT (which is stored within the intercepted context of the message), precisely defines which of the data fields can be sent safely to the client, and to what level of detail. 

Data, including personal data, can reveal different types of information, which have varying levels of dependence on coherence (e.g., ZIP~codes or phone numbers lose different amounts of information when the last digit is removed). This calls for the effective application of different data minimization techniques. The interceptor determines not only whether a data field can exist in the outgoing response message, but also controls its level of detail. Ultimately, data minimization is governed by the content of the PDP's decision and the contents of the response message.

Data minimization techniques for this proof of concept prototype include generalization, noising, reduction, and complete suppression of values. Since we published our contribution as an open-source project, implementing additional techniques is encouraged \textit{(FR3)}. We provide a detailed description of the implementation of each method in Sect.~\ref{PEP}.

\section{Implementation}\label{section:implementation}

As introduced in section \ref{section:approach}, the reusable component of our approach is divided along the XACML component structure. First, we will describe the practical implementation of the PAP and PDP, followed by the implementation of the gRPC interceptor which performs the PEP.

\subsection{Policy Administration and Decision}\label{PAP}

To achieve our goal of providing a solution that enables privacy techniques, such as automated (purpose-driven) policy enforcement and data minimization, we propose a structured machine-readable policy format. To be fully compatible with JWTs, we define a JSON-based policy.\footnote{\href{https://github.com/PrivacyEngineering/purpl-jwt-go-rsa}{github.com/PrivacyEngineering/purpl-jwt-go-rsa}} First, it comprises a list of service objects. Each service object can have a list of purpose objects in a flat structure, which distinguishes the data fields in the following categories: \texttt{allowed}, \texttt{generalized}, \texttt{noised}, and \texttt{reduced}. The data fields in these category objects may include a parameter to determine the applied minimization techniques more precisely. These options will be explained in section \ref{PEP}.

Furthermore, we implement the policy to be the single source of truth for the inter-service communication of the whole system. However, it should still be possible to use multiple policies in a system (up to one policy per service) since organization structures may not allow access to a system-wide policy. Nevertheless, having multiple policies necessitates well-defined policy management %
for avoiding bypassing or the circumvention of policy decisions. %
Our solution is suitable for individualization since the claims in the generated JWT are system-wide and their origin does not affect the interceptor behavior.

We implement our interceptor using the Go programming language, as being a natural high-performance fit to gRPC. Our implemented module generates the JWT based on five parameters: The \texttt{service name}, \texttt{purpose}, and \texttt{policy path} are used to get the corresponding data fields from the policy and parse them into the JWT claims. The \texttt{key} path is used to retrieve the provided Rivest–Shamir–Adleman (RSA) private key %
and to sign the JWT. Finally, the \texttt{expirationInHours} parameter sets the expiration of the JWT. %

In addition to our RSA-based module, we implemented the same functionality for a %
Elliptic Curve Digital Signature Algorithm (ECDSA) private key and published it as a separate Go module\footnote{\href{https://github.com/PrivacyEngineering/purpl-jwt-go-ecdsa}{github.com/PrivacyEngineering/purpl-jwt-go-ecdsa}}. We decided to implement these two algorithms since both are broadly used and supported by the module to handle JWTs.\footnote{\href{https://github.com/golang-jwt/jwt}{github.com/golang-jwt/jwt}}

Having the policy administration and decision separated from the policy enforcement, we abstract the token generation from the interceptor and, therefore, decrease its overhead. For the interested reader, we provide a simple overhead comparison of both approaches.\footnote{\href{https://github.com/PrivacyEngineering/purpl-naive-approach}{github.com/PrivacyEngineering/purpl-naive-approach}} %

\subsection{Policy Enforcement}\label{PEP}

\begin{figure}[!t]

\noindent\begin{minipage}[b]{.48\textwidth} %
    \begin{algorithm}[H] %
        \caption{Schematic description of the gRPC interceptor.}\label{alg:interceptor}
        \begin{algorithmic}
            \scriptsize
            \Require $JWT.expiration > \text{time.now}$
            \Require $JWT.signature = \text{valid}$
            \State $policy \gets JWT.policy$
            \ForAll{fields in message}
                \If{$field \in policy.allowed$}
                    \State $pass$
                \ElsIf{$field \in policy.minimized$}
                    \State $message.field \gets \text{minimize}(field)$
                \Else
                    \State $message.field \gets \text{suppress}(field)$
                \EndIf
            \EndFor
            \State \textbf{return} message
        \end{algorithmic}
    \end{algorithm}
\end{minipage}%
\hfill %
\begin{minipage}[b]{.48\textwidth} %
    \centering
    \includegraphics[width=\textwidth]{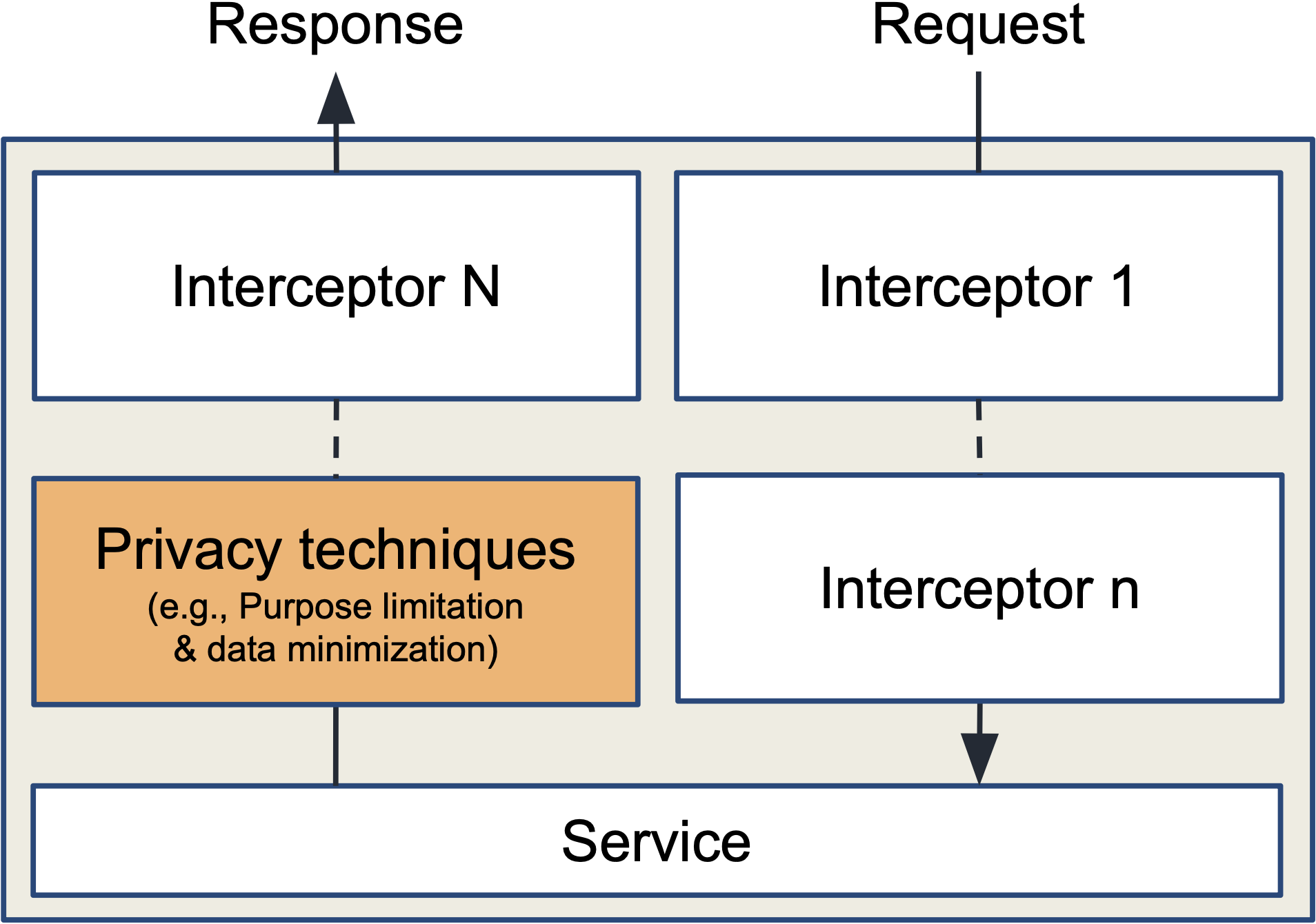}
    \captionof{figure}{gRPC interceptor chaining.} 
    \label{fig:interceptor}
\end{minipage}
\end{figure}

Whenever a response message is to be dispatched from the server, our interceptor,\footnote{\href{https://github.com/PrivacyEngineering/purpl}{github.com/PrivacyEngineering/purpl}} will be activated within the usual \texttt{grpc.NewServer()} function. The subsequent actions of the interceptor are as follows. Initially, the JWT is subjected to origin and expiration time verification. Once these checks pass successfully, the client-specific access policy from the PDP (as described in Sect.~\ref{PAP}) is extracted from the JWT and stored in a struct. Concurrently, the data field names from the response are extracted and stored in a slice. Having isolated the client-specific policy from the PDP and the data field names from the response message, the interceptor then performs the critical privacy technique, such as a data minimization task. The pseudocode as seen in algorithm \ref{alg:interceptor} describes this workflow on a high level.

We intended this algorithm to introduce only reasonable performance overheads.
For each field name in the response message, the interceptor determines whether the field should remain unmodified, needs to be minimized, or must be completely suppressed. Our implementation is inspired by common privacy techniques. For instance, all four data minimization mechanisms can handle integers, floats, and strings. Yet, some mechanisms differ in functionality depending on the data type that needs to be minimized, as follows. %
\medskip

\textit{Suppression} of a data field leads to the maximum information loss while maintaining the initial data types. Numeric values, such as integers and floats, are suppressed to a $-1$, while a string value is suppressed to an empty string (if needed differently, this can be changed easily). This guarantees the intended information loss while maintaining compatibility within the receiving programs, should they require the respective data types to be returned by the server. If, for example, the integer value of $42$ were to undergo suppression, the result would be $-1$. For compatibility reasons, the client would still receive an integer value for this data field, but every additional information would be lost in the minimization process. Here we point out that full suppression (i.e., removal of the data field entirely) is also technically possible in gRPC.   
\medskip

\textit{Generalization} of a data field leads to a reduction of the value precision. The information conveyed by the data should be neither lost nor altered completely, while still making the data less accurate. This mechanism is implemented for integers and floats by passing the respective value and a range parameter to the function. The range is defined in the JWT's policy and might change depending on the informational context. Assuming a data field \texttt{age} with the value $25$ were to undergo generalization with a range parameter of $10$, then the result would come out as $21$. $21$ representing, in this case, the age range from $21$ up until $30$. Respectively, $31$ represents the range $31 - 40$. In a different context, the parameter might change (e.g., \texttt{accountBalance:} $2.300$ with a parameter of $1000$ would return $2.001$). The chosen mapping ensures numbers larger than zero to always maintain this one property (but could be changed easily if needed).
Similarly to the numeric operations, invoking the generalization function for a string value in combination with a parameter will decrease the data's accuracy without altering it entirely. In these cases, the parameter specifies how many characters are to be returned. A \texttt{name: "Alice"} with parameter $1$ would thus be generalized to \texttt{name: "A"}.
\medskip

\textit{Noising} of a data field leads to an intended information loss, while maintaining a vague context of the initial data. This mechanism employs Google's differential privacy Go library.\footnote{\href{https://github.com/google/differential-privacy/tree/main/go}{github.com/google/differential-privacy/tree/main/go}} Our noising function can apply either \texttt{Laplace} or \texttt{Gaussian} noise to an input value of type integer or float. Due to the probabilistic nature of the noising function, an input value would be returned in a pseudo-random fashion (e.g., \texttt{age:} $25$ could be returned as \texttt{age:} $45$ in one and as \texttt{age:} $7$ in a subsequent function call).
We implement the handling of string-type values for robustness, while invoking the noising function will here lead to its suppression (as described above). For actually achieving  differential privacy, scenario-specific extensions would need to be implemented additionally. %

\medskip

\textit{Reduction} of data fields follows a similar idea as the generalization mechanism, but offers greater flexibility.
Reducing an integer or float value requires passing of a parameter value, which will then be used as a divisor in a simple division calculation (e.g., a \texttt{houseNumber:}~$135$ with parameter $10$ will be returned as \texttt{houseNumber:}~$13$, while a parameter $5$ would lead to a \texttt{houseNumber:}~$27$, due to the nature of integers).
A reduction of string-typed values, on the other hand, follows the same mechanism as the aforementioned generalization of strings. A use case could be the reduction of a \texttt{ZIP code} data field to its first four digits (e.g., \texttt{10623} to \texttt{1062}). Thus, not contradicting the initial information, while losing accuracy through broadening the geographical scope.  

\medskip
Ultimately, any field that requires minimization will be altered using the functions mentioned above. The output of the respective minimization function is used to overwrite the original message content with the \texttt{ProtoReflect().Set()} function. %
We support \texttt{protobuf} $v1.5.0$ to be used for inter-service communication. %
Once all message fields have been minimized according to the policy, the modified message handler, and consequently the message itself, will be returned and transmitted to its intended destination service.

\subsection{Usage and Configuration Mechanism}
To integrate the two reusable components, both referenced Go modules need to be included in the respective microservice, following our documentation. %
After successful integration, every privacy technique can be defined in the access policy and enforced through the gRPC interceptor. The interceptor can also be chained with other existing interceptors, as shown in Fig.~\ref{fig:interceptor}.
Within a service policy for a defined purpose, the data field names can be listed in either the \texttt{allowed} object or in one of the minimization objects. Fields listed in \texttt{generalized}, \texttt{noised}, or \texttt{reduced} require the specification of a parameter, as described in the previous section. Not documenting a field in one of the four objects will lead to its suppression, in case the data field appears in a response message.

\section{Preliminary Performance Evaluation}
\label{section:evaluation}

In the following, we summarize our preliminary performance assessment.

\textit{Scenario:} \label{use-case}
We assume a food delivery platform as a use case. Such services are widely utilized across the globe and inherently deal with personal information, such as address or payment information, detailed purchase histories, or demographic data.
In real settings, the collected information is actively shared with other parties for multiple different purposes.
For example,
contact information will have to be shared with the restaurants that prepare the food and the riders delivering the food, while %
demographic or device data will be used for internal research, technical, or marketing purposes. It would be disastrous if the marketing department could access banking information, without a valid legal basis under Art.~6~GDPR. %
Data minimization is also an important aspect, as the marketing department might want to use demographic data, such as age and place of residence, for a more focused marketing campaign. However, there is no need for detailed information, because the generalization of the data, e.g., an age range or the district of the residence, can already yield the needed results. %

To represent such a usecase, we modified the Online Boutique,\footnote{\href{https://github.com/GoogleCloudPlatform/microservices-demo}{github.com/GoogleCloudPlatform/microservices-demo}} which is a sample open-source microservice-based e-commerce application, initially provided by the Google Cloud Platform developers. %
The inter-service communication is gRPC-based, so we implemented our use case by expanding the architecture with an additional microservice, namely the \texttt{trackingservice}. %
It requests personal data like the address, name, and contact information to calculate the shortest route to the destination and displays the information, as seen by a potential delivery person, to the customer. We provided the \texttt{trackingservice} with multiple different purpose specifications, and each of them has a varying degree of allowed or restricted access to the requested information. %
For the following evaluation, we deployed the application to the Google Kubernetes Engine (GKE). 
Further details and instructions to reproducible the experiments are provided via Github.\footnote{\href{https://github.com/PrivacyEngineering/purpl-pizza-boutique/tree/main/terraform-gcp}{github.com/PrivacyEngineering/purpl-pizza-boutique/tree/main/terraform-gcp}}

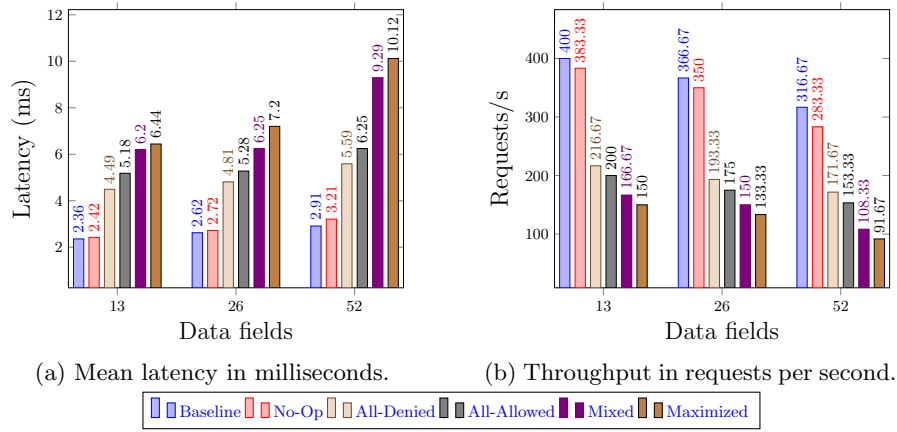
\begin{figure}[t]
\centering
\begin{subfigure}[h]{0.45\textwidth}
\begin{tikzpicture}[scale=0.65]
\begin{axis}[
    x tick label style={
        /pgf/number format/1000 sep=},
    xlabel=Data fields,
    ylabel=Latency (ms),
    label style={font=\Large},
    enlargelimits={abs=1cm},
    legend style={at={(0.5,1.03)}, anchor=south,legend columns=-1, legend to name={mylegend},nodes={scale=0.7}},
    legend entries={Baseline, No-Op, All-Denied, All-Allowed, Mixed, Maximized},
    ybar=3pt, %
    bar width=6pt, %
    symbolic x coords={13,26,52},
    xtick=data,
    nodes near coords,
    every node near coord/.append style={rotate=90, anchor=west},
]
\addplot %
    coordinates {(13, 2.356 ) (26, 2.623) (52, 2.912) };
\addplot %
    coordinates {(13, 2.423 ) (26, 2.723) (52, 3.209) };
\addplot %
    coordinates {(13, 4.4917 ) (26, 4.81) (52, 5.588) };
\addplot %
    coordinates {(13, 5.182 ) (26, 5.277) (52, 6.248) };
\addplot %
    coordinates {(13, 6.2009 ) (26, 6.2451) (52, 9.294) };
\addplot[fill=brown] %
    coordinates {(13, 6.4416 ) (26, 7.2) (52, 10.121)};

\end{axis}

\end{tikzpicture}
\caption{Mean latency in milliseconds.} 
\label{fig:latency}
\end{subfigure}%
~
\qquad %
\begin{subfigure}[h]{0.45\textwidth}
\begin{tikzpicture}[scale=0.65]
\begin{axis}[
    x tick label style={
        /pgf/number format/1000 sep=},
    xlabel=Data fields,
    ylabel=Requests/s,
    label style={font=\Large},
    enlargelimits={abs=1cm},
    ybar=3pt, %
    bar width=6pt, %
    symbolic x coords={13,26,52},
    xtick=data,
    nodes near coords,
    every node near coord/.append style={rotate=90, anchor=west},
]
\addplot %
    coordinates {(13, 400) (26, 366.6666667) (52, 316.6666667) };
\addplot %
    coordinates {(13, 383.3333333 ) (26, 350) (52, 283.3333333) };
\addplot %
    coordinates {(13, 216.6666667 ) (26, 193.3333333) (52, 171.6666667) };
\addplot %
    coordinates {(13, 200 ) (26, 175) (52, 153.3333333) };
\addplot %
    coordinates {(13, 166.6666667 ) (26, 150) (52, 108.3333333) };
\addplot[fill=brown] %
    coordinates {(13, 150 ) (26, 133.3333333) (52, 91.66666667)};
\end{axis}
\end{tikzpicture}

\caption{Throughput in requests per second.} 
\label{fig:thtoughput}
\end{subfigure}
\ref{mylegend}
\caption{Performance overheads for 3 different message sizes and 6 degrees of operational complexity.}
\end{figure}

We begin to assess the performance overhead generated by our purpose limiter technique. %
The experiments consist of a load generator imitating 10 users sending concurrent gRPC requests to the \texttt{trackingservice}. Each experiment iteration lasts ten minutes. The number of data fields in the response message and the kind of data minimization method are modified at every iteration. We assume that both the different minimization methods and the overall length of the response message can influence the performance of the gRPC interceptor. Fig.~\ref{fig:latency} depicts the measured latency in milliseconds, while Fig.~\ref{fig:thtoughput} illustrates the measured throughput. In both figures, the \textit{Baseline} represents communication without an interceptor, \textit{No-Op} communication with an interceptor performing no operations, \textit{All-Denied} suppression of all data fields, for \textit{All-Allowed} every data field is allowed to pass the interceptor without data minimization applied, \textit{Mixed} a variety of allowed fields and data minimization methods invoked, and \textit{Maximized} is the minimization methods on all data fields present.

\textit{Latency:}
The mere use of an interceptor (see Fig.~\ref{fig:latency}), even without performing any additional operations (\textit{No-Op}), always made a measurable impact compared to the \textit{Baseline}. We observe that the fastest performing functionality of our purpose limiter is the \textit{All-Denied} scenario with an average increase of 88\%. At the same time, the \textit{All-Allowed} follows with an average increase of 108\%. %
More complex data minimization techniques being applied, as the \textit{Mixed} or \textit{Maximized}  %
cases, show increases up to 200\% compared to the \textit{Baseline}. %
Increased amounts of fields in a request lead to increased latency. However, the increase is within a reasonable margin considering that the number of fields has been increased up to 4-fold, while the measured latency of our slowest-performing minimization technique has reached a 1.57-fold increase.

\textit{Throughput:} Fig.~\ref{fig:thtoughput} shows the measured throughput of our requests with 13, 26, and 52 data fields respectively, comparing varying degrees of minimization techniques. %
The Baseline performs the best, while the No-Op follows closely behind.
The loss of throughput is noticeable even with the fastest-performing \textit{All-Denied} scenario with an average loss of 47\% in throughput. %
The throughput decreases significantly with the amount of fields that need to be minimized. Increasing the number of data fields further also leads to a decrease in throughput, also for the \textit{Baseline}. Note, the relative loss of throughput is much smaller for the \textit{Baseline} than it is for scenarios that utilize many of the minimization techniques. %

Considering the amount of added computational complexity to an otherwise performance-optimized communication framework, such as gRPC, the measured latency and throughput fall into a reasonable range \textit{(NFR1)}. %
Additionally, the evaluations show a latency difference between the highest and lowest performing data minimization scenarios (\textit{All-Denied} and \textit{Maximized}) from 43\% to 81\%, while the throughput difference spans from 44\% to 87\%. Therefore, these findings suggest that the choice of data minimization scenario can significantly impact both latency and throughput, with potential variations. %
Albeit, advanced data minimization mechanisms will always be resource-intensive due to their computational complexity and the inherent need to explicitly handle single data fields. %
On the other hand, the relative overhead generated by our solution would probably decrease as soon as the corresponding microservice system itself increases in complexity.

\section{Limitations, Future Work and Conclusion
}\label{section:conclusion}

Given the nature of this work as a prototype, some limitations remain. First, %
when including the two Go modules in an application, actual secret management needs to be handled by the developers. For demonstration purposes, this aspect was excluded. %
Thus, public and private key generation should not be incorporated directly within the development environment. %

Further, the implementation of advanced purpose-based access control, including tree or graph structures of allowed/prohibited intended purposes, downstream usage policies, or transformation functions, seems a promising path for future work \cite{yappl}. The current prototype is limited by simple purpose specifications and does not yet fully implement the advances in this field \cite{wolf2021messaging}. Nevertheless, our general approach paves the way for such extensions.

Secondly, we propose to implement the PEP component as a \texttt{Stream\-Inter\-ceptor}. This would cover a second possible communication method, apart from unary interceptors, offered by gRPC, thus making the adaptation in existing microservice applications more likely. 
Moreover, %
the handling of further data types, apart from the ones described herein, %
such as complex objects, should be addressed. %
Lastly, the set of data minimization or masking methods should be extended to include as many options as possible (e.g., different hashing algorithms, actual differential privacy, $k$ anonymity of sets, etc.). %

Apart from additional features, further performance assessments could be conducted. The impact of policy size, for example, has not been measured yet. Regarding our assumption that validation using a JWT (generated from a tailor-made policy that only contains the necessary accepted and restricted data fields) might perform better than a JWT that contains many different purposes and fields that are not relevant for a respective service. Further measurements should also be accompanied by performance optimizations of the reusable components.

Regardless of the mentioned limitations and future work, we presented the first reusable approach that combines privacy techniques, such as data minimization and purpose limitation, natively into the gRPC communication framework. To illustrate the wide applicability of our contribution, we integrated our Go modules into an exemplary food delivery application. The observed performance overhead generated by our contribution is deemed reasonable. %
Ultimately, in the broader context of technical as well as legal privacy requirements, the importance of such technical contributions is evident. We addressed performance and implementation costs as two of the key factors in deciding whether data controllers are likely to implement the approach to ensure data protection by design and by default.

\subsubsection*{Acknowledgements.}\label{section:acknowledgements}

\begin{small}
We thank Huaning Yang, who contributed to the initial implementation within the scope of a privacy engineering course at TU~Berlin. 
\end{small}

\bibliographystyle{splncs04}
\bibliography{sn-bibliography}%

\end{document}